\documentclass[usenatbib]{mnras}
\usepackage[T1]{fontenc}
\usepackage{ae,aecompl,color}
\usepackage{graphicx} 
\usepackage{amsmath}  
\usepackage{amssymb}  
\usepackage{siunitx}  
\usepackage{widetext} 
\usepackage{cancel}

\newcommand{\mach}{\ensuremath{\mathcal{M}}}
\newcommand{\sh}{\mathrm{sh}}
\newcommand{\thm}{\mathrm{th}}
\newcommand{\cen}{\mathrm{cen}}
\newcommand{\esc}{\mathrm{esc}}
\newcommand{\tb}{\mathrm{turb}}
\newcommand{\total}{\mathrm{d}}
\newcommand{\crit}{\mathrm{crit}}
\newcommand{\eff}{\mathrm{eff}}
\newcommand{\msu}{Michigan State University, East Lansing, MI 48824, USA}

\title[Antesonic Condition]{The Antesonic Condition for the Explosion of Core-Collapse Supernovae II: Rotation and Turbulence}
\pubyear{2020}
\author[M. J. Raives et al.]{Matthias J. Raives$^{1,2}$, 
                             Todd A. Thompson$^{1,2}$, \&
                             Sean M. Couch$^{3,4,5,6}$
                             \\
                             ${}^{1}$Department of Astronomy, The Ohio State University, Columbus, OH 43210, USA\\
                             ${}^{2}$Center for Cosmology and Astroparticle Physics, The Ohio State University, Columbus, OH 43210, USA\\
                             ${}^{3}$Department of Physics and Astronomy, \msu\\
                             ${}^{4}$Department of Computational Mathematics, Science, and Engineering, \msu\\
                             ${}^{5}$National Superconducting Cyclotron Laboratory, \msu\\
                             ${}^{6}$Joint Institute for Nuclear Astrophysics-Center for the Evolution of the Elements, \msu
                             }
\begin{document}

    \label{firstpage}
    \pagerange{\pageref{firstpage}--\pageref{lastpage}}
    \maketitle
    
    \begin{abstract}
        In the problem of steady free-fall onto a standing shockwave around a central mass, the ``antesonic" condition limits the regime of stable accretion to $c_T^2/v_\mathrm{esc}^2\leq3/16$, where $c_T$ is the isothermal sound speed in the subsonic post-shock flow, and $v_\mathrm{esc}$ is the escape velocity at the shock radius. Above this limit, it is impossible to satisfy both the Euler equation and the shock jump conditions, and the system transitions to a wind. This physics explains the existence of a critical neutrino luminosity in steady-state 
        models of accretion in the context of core-collapse supernovae. Here, we extend the antesonic condition to flows with rotation and turbulence using a simple one-dimensional formalism. 
        Both effects decrease the critical post-shock sound speed required for explosion. While quite rapid rotation is required for a significant change to the critical condition, we show that the level of turbulence typically achieved in 
        supernova simulations can greatly impact the critical value of $c_T^2/v_\mathrm{esc}^2$. 
        A core angular velocity corresponding to a millisecond rotation period after contraction of the proto-neutron star results in only a $\sim\!5$ per-cent reduction of the critical curve. In contrast, near-sonic turbulence with specific turbulent kinetic energy $K/c_T^2=0.5-1$, leads to a decrease in the critical value of $c_T^2/v_{\rm esc}^2$ by $\sim\!20-40$ per-cent.
        This analysis provides a framework for understanding the role of post-shock turbulence in instigating explosions in models that would otherwise fail and helps explain why multi-dimensional simulations explode more easily than their one-dimensional counterparts.
    \end{abstract}
    
    \section{Introduction}
        
        A core-collapse supernova is initiated when the iron core of a massive star begins to collapse under its own gravity, beginning a runaway accretion process that is only halted when the core reaches nuclear density.  The sudden stiffening of the equation of state launches a shock wave into the accreting fluid, which loses energy as it moves outward, eventually stalling at a radius of a few hundred kilometers.  The central problem in supernova theory is explaining under what conditions, and by what mechanism, the shock is revived to produce a successful supernova.
        
        \citet{Burrows1993} quantitatively explored the concept of the critical condition for explosion. They showed that a spherically-symmetric accretion flow onto a standing accretion shock, subject to optically thin neutrino heating and cooling, only has a steady-state solution below a critical core neutrino luminosity $L_{\nu,\mathrm{crit}}$, which is a function of the mass accretion rate $\dot{M}$, as well as the proto-neutron star (PNS) mass $M$ and radius $R$.  Since then, numerical studies of the critical condition in time-dependent simulations have shown that the critical neutrino luminosity at a given $\dot{M}$ is generally lower in 2D (axisymmetry) and 3D than in 1D (spherical symmetry) \citep{Murphy2008,Murphy2017}, though higher in 3D than in 2D \citep{Couch2013b,Takiwaki2014}. Other simulations have investigated the critical curve with rotation \citep{Iwakami2014}, finding that rotation can drive a model to explosion even if the neutrino luminosity is below the ``critical'' value suggested by non-rotating simulations \citep{Yamasaki2005}. Work has also been done on explosion criteria that more directly ties the success or failure of explosion in simulations to characteristics of the progenitor models \citep{Ugliano2012,Pejcha2015,Sukhbold2016}, including the core compactness \citep{OConnor2011} and the $M_4$ mass \citep{Ertl2016}.
        
        \citet{Pejcha2012} showed that a critical condition similar to that of \citet{Burrows1993} can be derived for an isothermal equation of state in the post-shock accretion flow, where the ratio of the isothermal sound speed $c_T$ to the escape velocity at the shock is the controlling parameter. Above a critical value of $c_T^2/v_{\rm esc}^2=3/16$, it is impossible to simultaneously satisfy the time-steady Euler equations and the shock-jump conditions (see \S\ref{sec:ante}). \cite{Pejcha2012} also showed that this physics determines the critical condition in the more complete problem of \citet{Burrows1993}, and that the antesonic condition is not identical to the heuristic condition that the timescale for advection of matter through the gain region $\tau_\mathrm{advection}$ equals the heating timescale $\tau_\mathrm{heating}$ (e.g., \citealt{Thompson2004}). \citet{Pejcha2012} hypothesized that, in a time-dependent simulation, an accretion flow that exceeded the antesonic condition would dynamically transition to a thermally-driven wind. This predicted behavior was verified in time-dependent simulations by \citet{Gabay2015,Raives2018}.
       
       The applicability of the antesonic condition to the less idealized accretion flows present in full-physics multi-dimensional simulations of core-collapse supernovae with turbulence, convection, and other instabilities is unclear. In this paper, we take a step toward that understanding by deriving the antesonic condition for the idealized isothermal post-shock model flow with 1D prescriptions for turbulence and rotation. Importantly, our analysis shows that the critical condition is significantly affected by the near-sonic turbulence that occurs generically in multi-dimensional simulations, and thus provides a way to interpret many previous results. In contrast, in our model problem relatively extreme rotation rates producing millisecond rotation periods for the fully-contracted neutron star are required for the critical condition to be affected during the stalled-shock phase.  
       
       While previous works have investigated the effects of rotation \citep{Yamasaki2005} and turbulence \citep{Yamasaki2007,Mabanta2018,Couch2020} on the critical curve, here we are able to provide an analytic criterion for explosion that can be used to understand why some simulations explode, while others fail. We explicitly show how rotation and turbulence decrease the critical post-shock sound speed required for explosion.  In \S\ref{sec:ante}, we present the isothermal derivation from  \citet{Pejcha2012}. We then add rotation (\S\ref{sec:rotation}), both isotropic and anisotropic  turbulence (\S\ref{sec:turbulence}), and discuss a generalized critical condition allowing for rotation and turbulence (\S\ref{sec:fullante}).  In \S\ref{sec:discussion}, we propose a framework for understanding further generalizations to the antesonic condition (\S\ref{sec:ansatz}), as well as a comparison to both parameterized 1D (\S\ref{sec:couch}) and full-physics, multi-dimensional simulations (\S\ref{sec:multidim}).
    
    \section{The Antesonic Condition}\label{sec:ante}
        
    For context, we derive the simplest form of the antesonic condition, following \citet{Pejcha2012}.  An isothermal fluid undergoing steady-state shocked accretion in a point-mass gravitational potential must simultaneously satisfy both the Euler equations,
        \begin{align}
            \frac{1}{\rho}\frac{\total\rho}{\total r} &= -\frac{2}{r} - \frac{1}{v_r}\frac{\total v_r}{\total r}\label{eq:continuity}\\
            v_r\frac{\total v_r}{\total r} &= -\frac{1}{\rho}\frac{\total P}{\total r} - \frac{{GM}}{r^2}\label{eq:momentum},
        \end{align}
        and the shock-jump conditions,
        \begin{align}
            \rho_{1}v_{1} &= \rho_{2}v_{2},\\
            \rho_{1}v_{1}^2 + P_{1} &= \rho_{2}v_{2}^2 + P_{2},
        \end{align}
        as well as an equation of state (EOS).  For the purposes of this paper, we limit ourselves to an isothermal EOS of the form
        \begin{equation}
            P = c_T^2\rho
        \end{equation}
        in the post-shock fluid for simplicity. An exact antesonic condition can also be derived for a polytropic equation of state \citep{Raives2018}, and an antesonic condition also  holds for more general equations of state with heating and cooling \citep{Pejcha2012,Pejcha2015}. 
        
        In spherical symmetry, the Euler equations assume the dimensionless form
        \begin{equation}
            \left(\mach - \frac{1}{\mach}\right)\frac{\total\mach}{\total\xi_\thm} = \frac{2}{\xi_\thm} - \frac{1}{2\xi^2_{\thm}},
            \label{eq:euler}
        \end{equation}
        where $\mach=v_r/c_T$ is the Mach number of the fluid, and
        \begin{equation}
            \xi_{\thm}\equiv \frac{c_T^2r}{2{GM}} = \frac{c_T^2}{v_\esc^2}
        \end{equation}
        is the thermal ``antesonic" ratio, so named because this occurs at smaller $r$ than the sonic point in the classic Parker wind problem. Although in previous papers \citep{Pejcha2012,Pejcha2015,Raives2018}, we referred to this quantity as simply the ``antesonic ratio,'' without qualification, in this paper we attach the ``thermal'' prefix to distinguish it from the similar dimensionless ratios that are important to the additional physics considered below.
        
        \begin{figure}
            \centering
            \includegraphics[width=\linewidth]{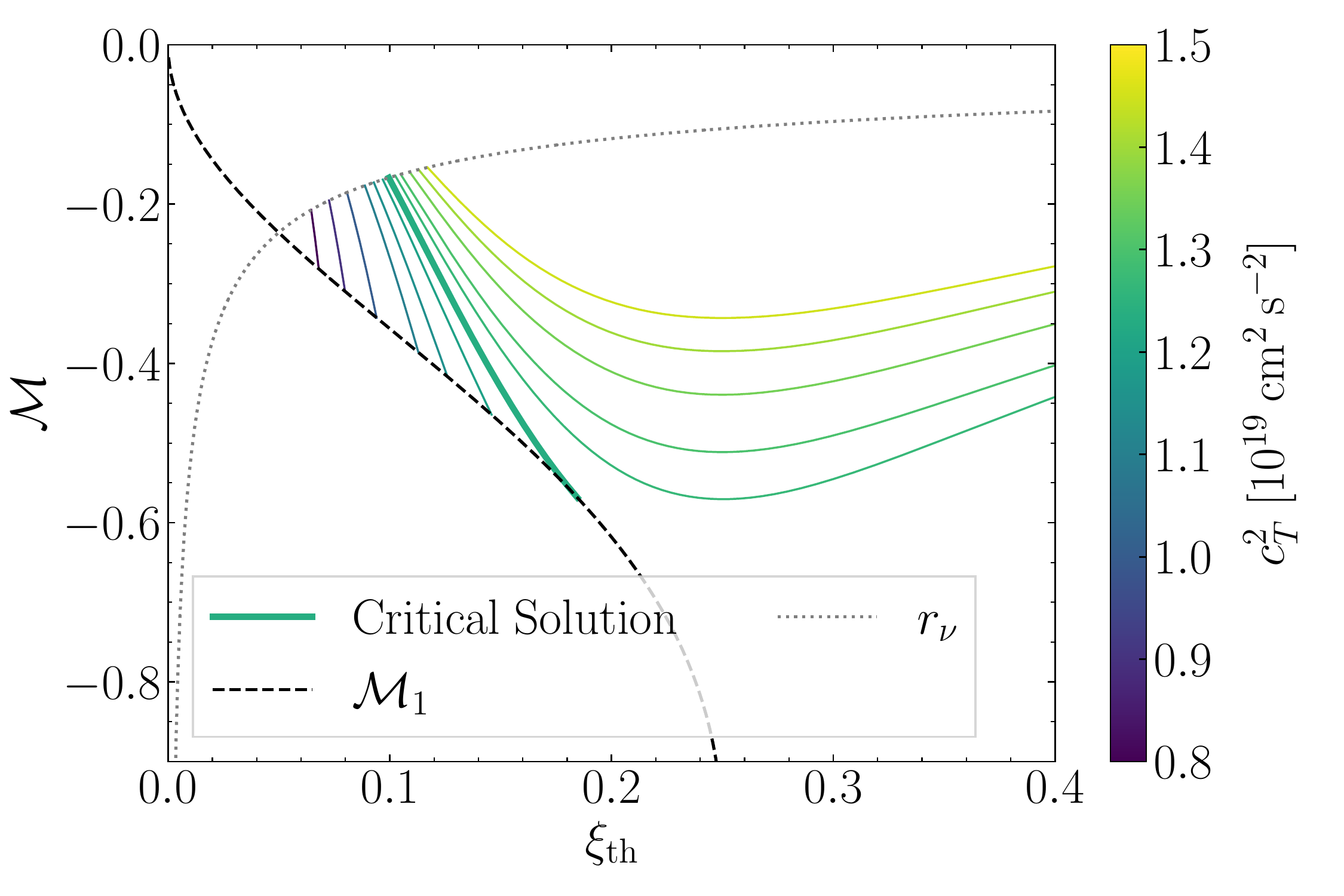}
            \caption{A graphical representation of the antesonic condition.  Solutions to the Euler equations (solid lines) are specified by the sound speed $c_T$ (increasing from left to right).  They start at $r_\nu=\SI{30}{\km}$ (dotted grey line) and continue until they satisfy the boundary conditions specified by the shock-jump conditions (Equation~\ref{eq:shockjump}, black dashed line). Solutions with a sound speed larger than that of the critical solution (bolded; $\xi_{\rm th}=3/16$) never satisfy the shock-jump conditions, and are thus unphysical in this context because they cannot describe shocked accretion flow.  The solutions pictured are for $\rho_\nu=\SI{3e10}{\gram\per\cm\cubed}$, $M=\SI{1.4}{M_\odot}$, and $\dot{M}=\SI{1}{M_\odot\per\second}$.}
            \label{fig:antesonic}
        \end{figure}
        
        Assuming the fluid above the shock is undergoing pressure-less free-fall ($P_2=0,$ $v_2=-\sqrt{2GM/r}$) onto the shock, the shock-jump conditions can be used to specify the fluid velocity immediately downstream of the shock as a function of the shock radius:
        \begin{equation}
            \mach_{1} = \frac{1}{2}\left(-\xi_\thm^{-1/2} + \sqrt{\xi_\thm^{-1}-4}\right).
            \label{eq:shockjump}
        \end{equation}
        By substituting Equation~\eqref{eq:shockjump} into Equation~\eqref{eq:euler}, we find the value (or values) of $\xi_\thm$ (and thus, $c_T$) that corresponds to a solution where the velocity profile is tangent to the locus of all possible shock positions (as specified by Equation~\ref{eq:shockjump}).  
        
        The importance of this solution is shown graphically in Figure~\ref{fig:antesonic}.  We integrated Equation~\eqref{eq:euler} outward from the PNS surface $r_\nu=\SI{30}{\km}$ (dotted line),  for various sound speeds at fixed assumed value of $\dot{M}=\SI{1}{M_\odot\per\second}$ (solid colored lines). Equation~\eqref{eq:shockjump} for the shock-jump conditions is shown as the black dashed line. For small $c_T$, the accretion flows go from the PNS surface to the shock, where they transition to the presureless free-fall conditions assumed to derive the shock jump condition (for clarity, these upstream profiles are not shown). For high values of $c_T$ (green to yellow), the accretion solutions do not intersect the shock conditions, indicating it is impossible to satisfy the time-steady Euler equation and the shock-jump conditions simultaneously. The dark green line indicates the critical solution where the two curves are tangent to one another.  For the pure isothermal case, the critical value of $\xi_\thm$ is equal to \citep{Pejcha2012}\footnote{The equations are also solved by $\xi_\thm=0,$ but this solution is unphysical.}
        \begin{equation}
        \xi_{\thm,\crit}=\frac{3}{16}.
        \label{eq:pureante}
        \end{equation}
        For isothermal and polytropic fluids \citet{Pejcha2012,Raives2018}, the critical value of $\xi_\thm$ is always the maximum value of $\xi_\thm$ anywhere in the accretion flow, and this maximum always occurs at the shock. However, this is not necessarily the case for other equations of state or when considering additional physics. As shown by \citet{Pejcha2012}, for an equation of state that is coupled to neutrino heating and cooling, the maximum value of $c_s^2/v_\esc^2$ occurs near the ``gain radius,'' where neutrino heating balances cooling in the post-shock flow.
        
        \subsection{Rotation}\label{sec:rotation}
            We first consider the case of a non-turbulent fluid rotating with velocity $v_\phi(r)$, which experiences a centrifugal acceleration in the equatorial plane:
            \begin{equation}
                a_\cen = \frac{v_\phi^2}{r} = 2\frac{GM}{r^2}\xi_\cen,
            \end{equation}
            where
            \begin{equation}
                \xi_\cen\equiv\frac{v_\phi^2}{v_\esc^2},
                \label{eq:xicen}
            \end{equation}
            so that Equation~\eqref{eq:momentum} becomes
            \begin{equation}
                v_r\frac{\total v_r}{\total r} = -\frac{1}{\rho}\frac{\total P}{\total r} - \frac{GM}{r^2}(1-2\xi_\cen)
            \end{equation}
            In dimensionless form, (for comparison with Equation~\eqref{eq:euler}), we have
            \begin{equation}
                \left(\mach - \frac{1}{\mach}\right)\frac{\total\mach}{\total\xi_\thm} = \frac{2}{\xi_\thm} - \frac{1-2\xi_\cen}{2\xi_\thm^2}.\label{eq:euler_rot}
            \end{equation}
            As the rotational velocity is the same on either side of the shock, the presence of rotation does not change the shock-jump conditions.  Following the same procedure as for the non-rotating case (i.e., substituting Equation~\eqref{eq:shockjump} into Equation~\eqref{eq:euler_rot}), we obtain the following expression for the critical antesonic ratio:
            \begin{equation}
                \xi_{\thm,\crit} = \frac{1}{32}\left(3-16\xi_{\cen,\crit} + \sqrt{9+32\xi_{\cen,\crit}}\right).
                \label{eq:rotante}
            \end{equation}
            In the limit $\xi_{\cen,\crit}\ll 1$, Equation~\eqref{eq:rotante} reduces to
            \begin{equation}
                \xi_{\thm,\crit} + \frac{1}{3}\xi_{\cen,\crit} = \frac{3}{16}.
            \end{equation}
            In the presence of rotation, there is not one single critical antesonic ratio but rather, the antesonic condition specifies a relation between the thermal and centrifugal terms at the shock that defines the critical solution.
            
\begin{figure}
                \centering
                \includegraphics[width=\linewidth]{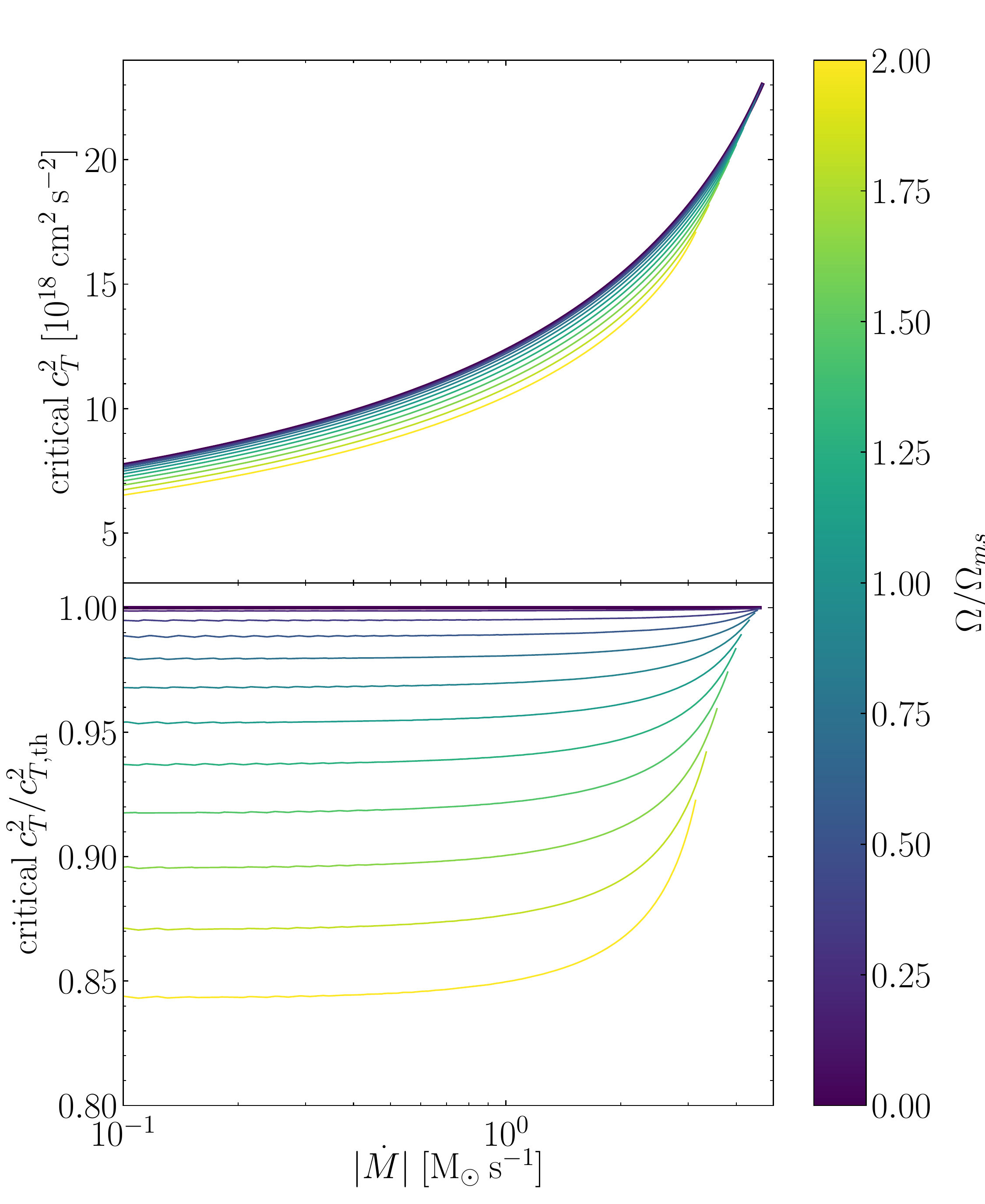}
                \caption{The critical sound speed $c_T$ for a given accretion rate $\dot{M}$, for varying core rotation rates $\Omega$, for the same PNS mass and radius as in Figure~\ref{fig:antesonic}.  $\Omega$ is shown in units of $\Omega_{ms}$, the rotation rate that produces a millisecond rotator after contraction (i,e., accounting for angular momentum conservation).  Faster rotation lowers the critical curve, i.e., reduces the critical $c_T^2$ for a given $\dot{M}$.  Faster rotation also lowers the maximum accretion rate that shocked accretion can support (above which, $R_\mathrm{sh}<R_\nu$).  The bottom panel shows the ratio of the critical curves for different rotation rates to the non-rotating curve.}
                \label{fig:rot}
            \end{figure}

            We follow the same procedure used to generate Figure \ref{fig:antesonic} to find the critical value of $c_T$ for many different assumed values of $\dot{M}$. For each $\dot{M}$, we search for the tangent of the flow profiles obtained from Equation~\eqref{eq:euler_rot} with the shock jump conditions. Because the value of the shock radius at the critical value of $c_T$ changes for each $\dot{M}$, the mapping between the antesonic condition given in Equation~\eqref{eq:pureante} or Equation~\eqref{eq:rotante} and $\dot{M}$ is not trivial. We then plot this ``critical curve'' showing the critical value of $c_T$ as a function of $\dot{M}$ in Figure~\ref{fig:rot}\footnote{In the full-physics case, the function instead gives the core neutrino luminosity $L_\nu$ \citep{Burrows1993,Pejcha2012}.}. In order to integrate Equation~\eqref{eq:euler_rot} and construct Figure~\ref{fig:rot}, we assume a rotation profile $v_\phi(r)$ corresponding to angular momentum conservation with $v_\phi \propto r^{-1}$.  We have also tested the case of a solid body rotation profile $(v_\phi\propto r)$ (see below).
            
            By allowing a second parameter to vary (in this case, the core rotation rate $\Omega$) we form a 2D critical surface (e.g., as in \citealt{Iwakami2014}). For clarity, we show individual slices of the critical surface at fixed values of $\Omega$.  For aid in interpreting the figure, we normalize the rotation rates by
            \begin{equation}
                \Omega_\mathrm{ms} = \frac{2\pi}{\SI{1}{\ms}}\left(\frac{\SI{10}{\km}}{r_\nu}\right)^2,
            \end{equation}
            the rotation rate that would lead to a neutron star with a period of \SI{1}{\milli\second} after the PNS cools and contracts to a radius of \SI{10}{\km}, assuming the angular momentum of the PNS is conserved during contraction, and that the core rotates as a solid body.  For this analysis, we assume a fixed $r_\nu=\SI{30}{\km}$.
            
            We find that, as expected, more rapidly rotating cases have lower critical curves, i.e., for a given mass accretion rate $\dot{M}$, the critical sound speed $c_T$ is smaller for large $\Omega$ than for small $\Omega$.    For $\Omega/\Omega_\mathrm{ms}=1$, the critical curve only decreases by $\sim5$ per-cent, indicating that, for these parameters and for this model problem, rotation has a relatively small effect on the critical curve. In terms of the critical antesonic ratios, this rotational velocity corresponds to $\xi_\cen = \num{7.08e-3}$ (for $r_\sh=\SI{150}{\km}$), which means the critical thermal antesonic ratio is $\xi_\thm \simeq 0.1851$, a 1 per-cent decrease.  When we instead assume solid-body rotation ($v_\phi(r)\propto r$) for the post-shock profile in solving Equation~\eqref{eq:euler_rot}, we find larger fractional decreases in the critical curve for the same core rotation $\Omega$ because the rotational velocity at the shock will be larger.
            
            We also note that each critical curve has a maximum mass accretion rate, which decreases with more rapid rotation.  For larger accretion rates and more rapid rotation, the shock radius of the critical solution is smaller; for sufficiently large $\dot{M}$ or $\Omega$, the shock radius will be smaller than the PNS radius, which is unphysical.  The ultimate physical limit on rotation is $\xi_\cen = \frac{1}{2}$, as above this value, Equation~\eqref{eq:rotante} implies that $\xi_\thm<0$, i.e., stable accretion is impossible at any sound speed.  The limit of $r_\sh>r_\nu$ provides a slightly more stringent limit, and other constraints, such as shock heating of the accreted material, or neutrino heating from the PNS core, will also tend to decrease the ``maximum'' rotation rate.  We intend to further explore such constraints in future work.
            
            While rotation rates $\Omega\sim\Omega_\mathrm{ms}$ may be relevant for core collapses that produce super-luminous magnetar-powered supernovae or gamma-ray bursts \citep{Thompson2004,Metzger2007,Kasen2010,Metzger2011}, they are much faster than expected for normal pulsars, which have $P\sim10-1000$\,ms (e.g., \citealt{FaucherGiguere2006}). Note that for clarity here we focus exclusively on the effect of the centrifugal force in the equatorial plane. In a more realistic model, especially at the rapid rotation rates needed for modification of the critical curve found here, other multi-dimensional effects become important (e.g., \citealt{Yamasaki2005}).
            
        \subsection{Turbulence}\label{sec:turbulence}
            We now consider the problem of a non-rotating, but turbulent flow beneath the shock.  We further assume that the flow is fully non-turbulent above the shock, i.e., that the turbulence is caused by neutrino heating and convection in the gain region \citep{Herant1994,Janka1996,Burrows1995} or the standing accretion shock instability \citep{Blondin2003,Couch2014,Fernandez2015}, as obtained by multi-dimensional supernova simulations.
            For a turbulent flow, we must separate each fluid variable $f$ into its background (i.e., Favre-averaged) component $\widetilde{f}$ and its turbulent component $f''$ \citep{Mocak2014}:
            \begin{equation}
                f = \widetilde{f} + f''.
            \end{equation}
            The Favre average is defined as:
            \begin{equation}
                \overline{\rho}\widetilde{f} = {\overline{\rho f}},\qquad \widetilde{f''} = 0,
            \end{equation}
            where $\overline{f}$ is the Reynolds average
            \begin{equation}
                \overline{f} = \frac{1}{\Delta T}\int_T f(t)\:\mathrm{d}t.
            \end{equation}
            The Favre average is thus a density-weighted Reynolds average.  While others \citep{Meakin2007,Arnett2009,Murphy2011} use the Reynolds average, we prefer the Favre average for reasons of mathematical convenience. Starting with the full, time-steady Euler equations, presented here in vector form:
            \begin{align}
                \nabla\cdot(\rho\mathbf{v}) &= 0\\
                \nabla\cdot(\rho\mathbf{v\otimes v}) + \nabla{P} - \rho\mathbf{g} &= 0
            \end{align}
            We then decompose $\mathbf{v}$ into its background and turbulent components, and take the Reynolds average of each equation.  This leaves us with
            \begin{align}
                \nabla\cdot(\overline{\rho}\widetilde{\mathbf{v}}) &= 0\\
                \overline{\rho}\widetilde{\mathbf{v}}\cdot\nabla\widetilde{\mathbf{v}} +  \nabla\cdot\widetilde{\mathbf{R}} + \nabla{\overline{P}} - \overline{\rho}\mathbf{g} &= 0.
            \end{align}{}
            Here, $\mathbf{R}$ is the Reynolds stress tensor, which characterizes the strength of turbulence.  Its components are given by
            \begin{equation}
                \widetilde{R_{ij}} = \overline{\rho}\widetilde{v_i''v_j''}.
            \end{equation}
            For our purposes, however, it is helpful to write things in terms of the specific turbulent kinetic energy:
            \begin{equation}
                K = \sum_i\frac{1}{2}\widetilde{v_i''v_i''} = \frac{1}{2\overline{\rho}}\sum_i\widetilde{R_{ii}}.
            \end{equation}
            We can characterize the degree of isotropy in the turbulence by comparing the kinetic energy in turbulent radial velocity to that in turbulent angular velocities:
            \begin{equation}
                \alpha \equiv \frac{\widetilde{R_{\theta\theta}} + \widetilde{R_{\phi\phi}}}{\widetilde{R_{rr}}}\label{eq:alpha}.
            \end{equation}
            A value of $\alpha=2$ corresponds to isotropic turbulence, while $\alpha=1$ corresponds to anisotropic turbulence with $\widetilde{R_{rr}}=\widetilde{R_{\theta\theta}} + \widetilde{R_{\phi\phi}}$, which is a better fit to the character of turbulence in the gain region \citep{Arnett2009,Murphy2013,Couch2015a,Radice2016}.  We can also make use of this parameter to define the total turbulent kinetic energy solely in terms of the radial Reynolds stress:
            \begin{equation}
                K = \frac{1+\alpha}{2\overline{\rho}}\widetilde{R_{rr}}
            \end{equation}
            Assuming an isothermal equation of state, spherical symmetry, and
            \begin{equation}
                \frac{\partial K}{\partial r} = 0,
                \label{kconstant}
            \end{equation}
            (an assumption required for a tractable analytic solution), we can write the Euler equations in the dimensionless form:
            \begin{equation}
                \left(\widetilde{\mach} - \widetilde{\mach}^{-1}\right)\frac{\partial\widetilde{\mach}}{\partial \xi_\eff} - \frac{2}{\xi_\eff} + \frac{\Xi_1}{2\xi_\eff^2} = 0,
            \end{equation}
            where
            \begin{align}
                \widetilde{\mach} &= \frac{\widetilde{v_r}}{c_\eff} \\
                c^2_\eff &= c_T^2 + \frac{2K}{1+\alpha} \\
                \xi_\eff &= \frac{c_\eff^2}{v_\esc^2} \\
                \Xi_1 &= 1 + 2\frac{2-\alpha}{1+\alpha}\xi_\tb \label{eq:Xi1}\\
                \xi_\tb &= \frac{2K}{v_\esc^2}.\label{eq:def}
            \end{align}
            
            The full derivation of the Favre-averaged Euler equations is presented in Appendix~\ref{app:Favre}.  We derive the antesonic condition under both isotropic and anisotropic turbulence in the following sub-sections.
            
            We note that many studies describe the strength of turbulence in the post-shock region by the turbulent Mach number (e.g., \citealt{Mueller2015}), which is nominally related to our turbulent antesonic ratio by
            \begin{equation}
                \xi_\tb \sim \xi_\thm \widetilde{\mach^2_\tb},
            \end{equation}
            where
            \begin{equation}
                \widetilde{\mach^2_\tb} = \frac{\widetilde{v_r''v_r''}}{c_T^2}
            \end{equation}
            However, care needs to be taken to account for the mix between radial and non-radial turbulent motions.  We return to this issue in Section \ref{sec:discussion}.

            \subsubsection{Isotropic Turbulence}
                For isotropic turbulence $(\alpha=2)$, the turbulent contribution to $\Xi_1$ vanishes.  That is:
                \begin{equation}
                    \left(\widetilde{\mach} - \widetilde{\mach}^{-1}\right)\frac{\partial\widetilde{\mach}}{\partial \xi_\eff} - \frac{2}{\xi_\eff} + \frac{1}{2\xi_\eff^2} = 0.
                \end{equation}
                This is identical to the pure isothermal case (Equation~\ref{eq:euler}) with the substitution $c_T\to c_\eff$.  A treatment of the shock-jump conditions similar to that of the pure isothermal case (see Appendix~\ref{app:shockjump} for a full derivation) gives us an equation like Equation~\eqref{eq:shockjump} with $c_T\to c_\eff$:
                \begin{equation}
                    \widetilde{\mach_1} = \frac{1}{2}\left(-\xi_\eff^{-1/2} + \sqrt{\xi_\eff^{-1} - 4}\right).\label{eq:FavreMach}
                \end{equation}
                Since both our Euler equation and shock-jump condition are the same as in the pure isothermal case, except with $\xi_\thm\to\xi_\eff$, the solution to this pair of equations -- i.e., the antesonic condition -- must be the same as the pure isothermal antesonic condition, except with $\xi_\thm\to\xi_\eff$, i.e.,
                \begin{equation}
                    \xi_\eff = \xi_\thm + \frac{1}{3}\xi_\tb = \frac{3}{16}.\label{eq:isoturbante}
                \end{equation}
                We note that this antesonic condition has a critical threshold at $\xi_\tb=\frac{9}{16}$, which corresponds to a turbulent kinetic energy of
                \begin{equation}
                    K_{\rm max} = \frac{9}{32}v_\esc^2\label{eq:isoturblimit}
                \end{equation}
                at the shock.  Above this threshold, a steady accretion flow cannot be maintained at \emph{any} finite value of the post-shock sound speed ($\xi_\thm>0$).  Like in the case with rotation, however, this limit is merely an upper bound on the maximum turbulent kinetic energy for which stable accretion is possible. Because the material is heated as it passes through the shock and then heated throughout the gain region, we expect the practical limit on $K$ to be smaller. This will be the subject of a future investigation.
                
            \subsubsection{Anisotropic Turbulence}
                Anisotropic turbulence with $\alpha=1$ is more characteristic of turbulence in the gain region than isotropic turbulence \citep{Arnett2009,Murphy2013,Couch2015a,Radice2016}.  With this prescription, the Euler equation becomes:
                \begin{equation}
                    \left(\widetilde{\mach} - \widetilde{\mach}^{-1}\right)\frac{\partial\widetilde{\mach}}{\partial \xi_\eff} - \frac{2}{\xi_\eff} + \frac{1+\xi_\tb}{2\xi_\eff^2} = 0.
                \end{equation}
                This equation is analogous to the rotational case, but with $\xi_\cen\to-\frac{1}{2}\xi_\tb$ and $c_T\to c_\eff$.  As the shock-jump conditions imply the same functional form of the post-shock Mach number as in the isotropic case (i.e., Equation~\eqref{eq:FavreMach}), this implies that our antesonic condition must be
                \begin{equation}
                    \xi_\thm + \frac{1}{2}\xi_\tb = \frac{1}{32}\left(3 + 8\xi_\tb + \sqrt{9 - 16\xi_\tb}\right). \label{eq:turbante}
                \end{equation}
                which can be directly compared to Equations~\eqref{eq:pureante} and \eqref{eq:rotante} for the pure isothermal antesonic condition and the antesonic condition with rotation, respectively.  We note that this equation has a root at $\xi_{\tb} = 1/2$, which corresponds to a maximum value of the turbulent kinetic energy of
                \begin{equation}
                    K_{\rm max} = \frac{1}{4}v_{\rm esc}^2
                    \label{eq:turblimit}
                \end{equation}
                at the shock -- similar to, though slightly smaller than, the similar limit for isotropic turbulence given in Equation~\eqref{eq:isoturblimit}. As in the isotropic case, we expect the practical limit on $K$ to be significantly less than this value because of shock heating and neutrino heating in the gain region, an issue we will return to in a future work.
                
                In the limit where $\xi_\tb\ll1$, Equation~\eqref{eq:turbante} has the approximate form
                \begin{equation}
                    \xi_\thm + \frac{1}{3}\xi_\tb = \frac{3}{16}.\label{eq:turbanteappr}
                \end{equation}
                
\begin{figure}
                \centering
                \includegraphics[width=\linewidth]{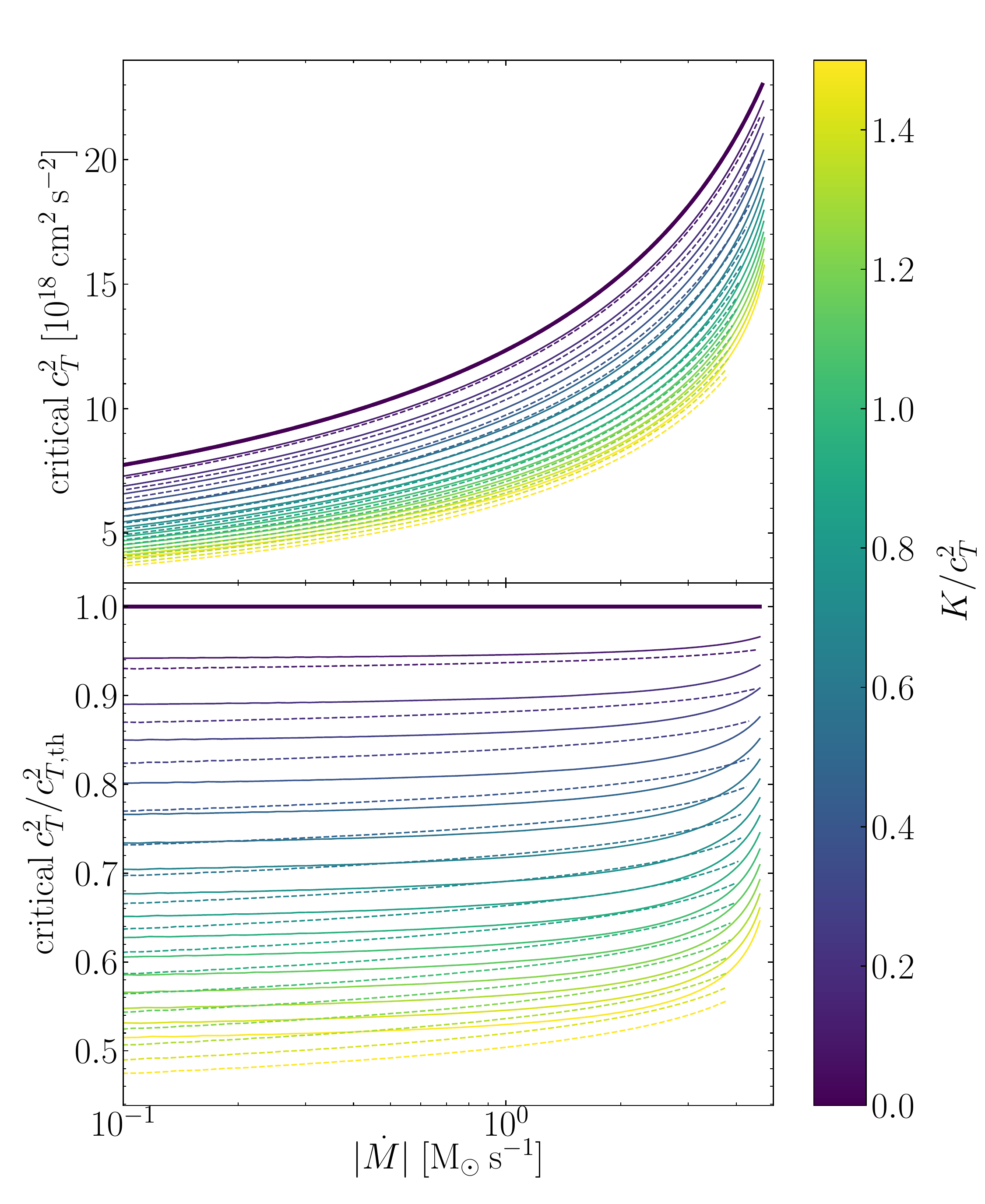}
                \caption{The critical curve for varying turbulent kinetic energies (shown here in units of $c_T^2$), for both isotropic (solid) and anisotropic (dashed) turbulence.  We see that stronger turbulence lowers the critical curve, and that anisotropic turbulence also has a lower critical curve than isotropic turbulence does.  Turbulence, both isotropic and anisotropic, has a larger effect on the critical curve at lower $\dot{M}$.}
                \label{fig:turb}
            \end{figure}

\begin{figure}
                \centering
                \includegraphics[width=\linewidth]{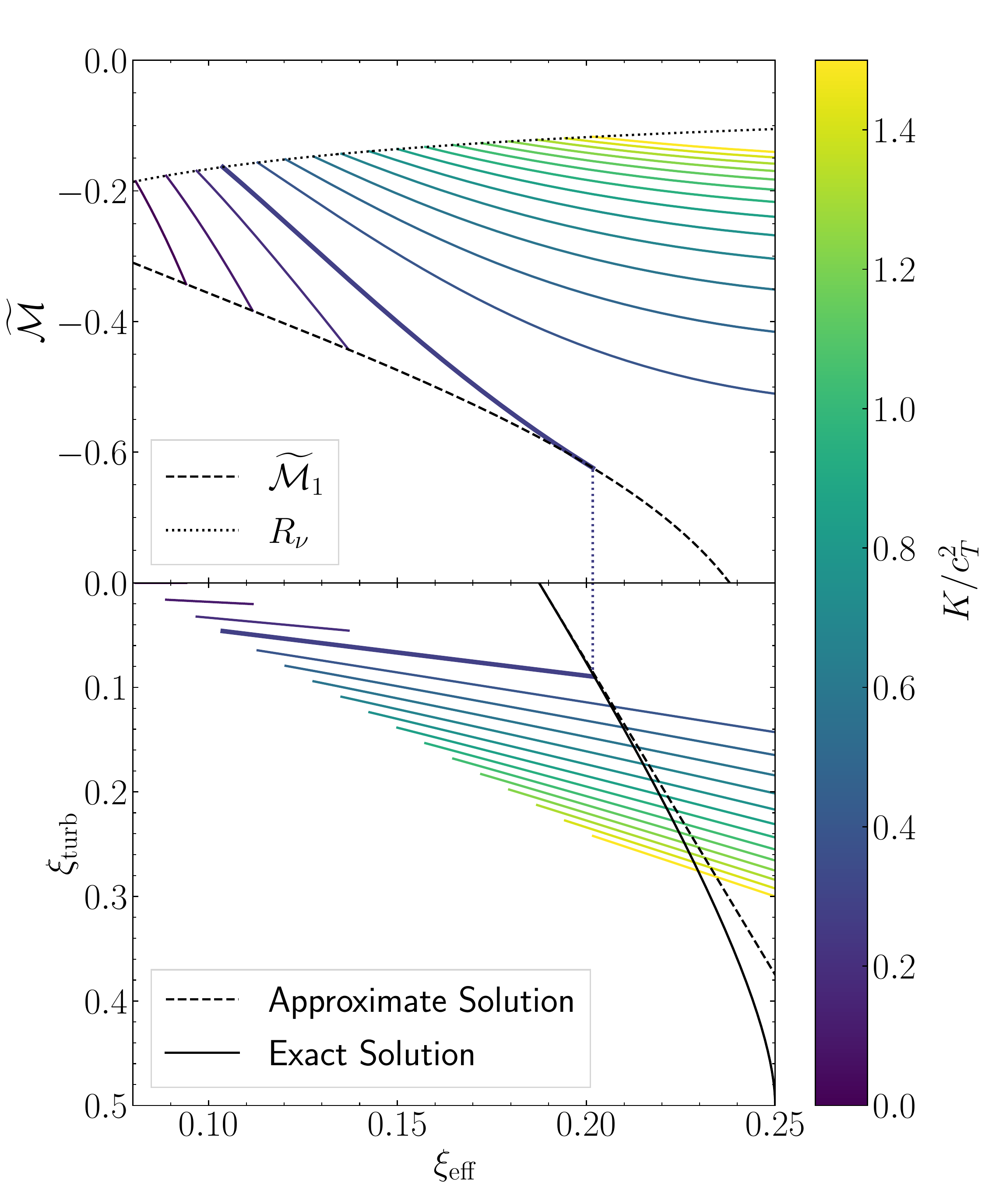}
                \caption{A visual demonstration of Equation~\eqref{eq:turbante}, analogous to Figure~\ref{fig:antesonic}.  Top: the Mach number $\widetilde{\mach}$ vs the effective antesonic ratio ${
                \xi_\eff=c_\eff^2 r/2GM=\xi_\thm+\frac{1}{2}\xi_\tb}$, with curves color-coded by the ratio of $K/c_T^2$ (colors correspond to the same values of $K/c_T^2$ as in Figure~\ref{fig:turb}).  Though each of these curves has the same $c_T^2=\SI{e19}{\cm\squared\per\second\squared}$ and ${\dot{M}=\SI{1.0}{M_\odot\per\second}}$, they start at different $\mach$ and $\xi_\eff$ because of the $K$ contribution to both those terms. Bottom: the turbulent antesonic ratio vs the effective antesonic ratio $\xi_\tb$ (note the reversed y-axis) for each solution shown in the top panel.  The solid and dashed lines correspond to Equation~\eqref{eq:turbante} and Equation~\eqref{eq:turbanteappr} (which is also the approximate critical curve for anisotropic turbulence), respectively.  We see that the critical solution, which is bolded in each panel, just grazes the curve describing the Mach number at the shock (Equation~\ref{eq:FavreMach}) at the same point where it touches the curve describing the antesonic condition.}
                \label{fig:turbAnte}
            \end{figure}

                We use the same procedure used in constructing Figure \ref{fig:rot} to calculate the critical curves for isotropic to anisotropic turbulence in Figure~\ref{fig:turb}. For many different values of $\dot{M}$, we determine the critical value of the sound speed above which accretion is impossible, for a range of values of $K$. We see that with increasing turbulent kinetic energy (larger $K/c_T^2$), the normalization of the critical curve decreases. We also see that, except at the highest mass accretion rates, the decrease in normalization is roughly constant with $\dot{M}$.  In Figure~\ref{fig:turbAnte}, we visually demonstrate the turbulent antesonic condition (Equation~\ref{eq:turbante}).  The figure is analogous to Figure~\ref{fig:antesonic}, except that we vary the ratio $K/c_T^2$ while keeping $c_T$ constant.  We find that, like with increasing $c_T$ in the case dominated by thermal pressure, increasing the turbulent kinetic energy moves the shock radius outwards until the critical solution is reached.
                
                Importantly, this exercise shows that for values of $K/c_T^2$ obtained in multi-dimensional supernova simulations (e.g., \citealt{Murphy2013,Couch2015a,Radice2016,Mabanta2018}), the fractional decrease in the critical curve at fixed $\dot{M}$ can be large. While simulations find a range of turbulent kinetic energies as a function of both radius and time in a given massive star progenitor, values of $K$ are often in the near-sonic range, with $K/c_T^2\sim1$, implying order-unity decrease to critical curve.

        \subsubsection{Power Law $K(r)$ Profiles}
        
            In our analysis, we assume that the turbulent kinetic energy density $K(r)$ is constant with radius (Equation \ref{kconstant}), as this assumption allows us to find a straightforward analytic solution to the problem.  However, simulations \citep{Radice2016} suggest a power law $K(r)\propto r^\beta$ might be more appropriate.  From that study in particular, we could approximate $K\propto r^3$ in the gain region.  The general effect of such a parameterization would be to have stronger turbulence near the shock than near the core.  Though we also note that turbulence enters the momentum equation as ${\partial (\overline{\rho}K)}/{\partial r}$ so that if $K$ increases with $r$, then this will dampen the effect of the turbulent pressure gradient.
            
            We tested numerical models with a power-law $K(r)$ and found that, compared to models with the same $K(r_\nu)$ but constant $K(r)$, power-law models with $\beta>0$ had larger shock radii, and larger values of $\xi_\tb$ and $\xi_\eff$ at the shock.  Furthermore, the critical solution had smaller $K(r_\nu)$, a smaller shock radius, and a larger $\xi_\tb$ and $\xi_\eff$ at the shock than the critical solution of the constant $K(r)$ model.  For the $\beta<0$ case, the reverse was true: for a given $K(r_\nu)$, the power law model had a smaller shock radius, and smaller $\xi_\tb$ and $\xi_\eff$ at the shock.  The critical solution occurred at larger $K(r_\nu)$, and had a larger shock radius, and smaller $\xi_\tb$ and $\xi_\eff$ at the shock.  However, although the critical solutions occur at different $\xi_\thm$ and $\xi_\tb$, we find that these values still respect the critical curve as defined in Equations \eqref{eq:isoturbante}~and~\eqref{eq:turbante}.
            
        \subsection{Combining Rotation and Turbulence}\label{sec:fullante}
            We can write the antesonic condition without specifying the exact nature of the turbulent balance (i.e., for arbitrary $\alpha$), while also including rotation.  This condition takes the form:
            \begin{equation}
                \xi_\eff = \frac{1}{32}\left(8\Xi_2 + \sqrt{25 - 16\Xi_2} - 5\right),\label{eq:rotandturb}
            \end{equation}
            where (see Equations~\ref{eq:xicen} and \ref{eq:Xi1}-\ref{eq:def})
            \begin{equation}
                \Xi_2 = \Xi_1 - 2\xi_\cen.
            \end{equation}
            In the limits $\xi_\cen\ll1,\;\xi_\tb\ll1$, this reduces to
            \begin{equation}
                \xi_\thm + \frac{1}{3}\xi_\cen + \frac{1}{3}\xi_\tb = \frac{3}{16}.\label{eq:firstorder}
            \end{equation}
            
        \section{Discussion}\label{sec:discussion}
            
        \subsection{An Antesonic ``Ansatz''}\label{sec:ansatz}
            Both rotation and turbulence add to the critical condition in a manner that is to first order quite straightforward and intuitive. A simple, but useful way to think about any additional forces that could be added to the Euler equation, is that they will add to the isothermal antesonic condition in a similar manner:
            \begin{equation}
                \xi_\th + \frac{1}{3}\xi_\cen + \frac{1}{3}\xi_\tb + \cdots + a_i\xi_i \approx \frac{3}{16},
                \label{eq:ansatz}
            \end{equation}
            where the $\xi_i$ term takes the form
            \begin{equation}
                \xi_i = \frac{v_{i}^2}{v_\esc^2},
            \end{equation}
            and $a_i$ is a coefficient, with potentially negative values for terms that correspond to forces acting in the same direction as gravity.  The $v^2_i$ here is the characteristic velocity (squared) for the force in question, where $v^2_\phi$ and $2K$ are the relevant velocities for the cases of rotation and turbulence, respectively. 
            
            This ``ansatz'' for the form of a generalized antesonic condition allows for simple estimates of how big certain terms might in principle need to be in order to affect the dynamics. For example, if we were to consider a simple radial magnetic pressure term
            \begin{equation}
                P_\mathrm{mag} = \frac{1}{2}\frac{B^2}{4\pi},
            \end{equation}
            (i.e., ignoring magnetic tension), the antesonic condition would contain the term
            \begin{equation}
                \xi_\mathrm{mag} = \frac{v_A^2}{v_\esc^2},\quad a_\mathrm{mag} = \frac{1}{2},
            \end{equation}
            in the simplest case where we assume $\partial v_A/\partial r = 0$. To significantly modify the critical curve, near-magnetosonic $v_A\sim c_T$ would be required, as might be expected from near-sonic turbulence (e.g., \citealt{Ostriker2001}). Recent studies \citep{Mueller2020} suggest that magnetic fields can play a central role in explosion for $E_\mathrm{mag}\sim\frac{1}{2}E_\tb$.  As with the case of turbulence considered above, other forms of $v_A(r)$ may lead to the same antesonic condition, but different shock radii for different normalizations of the post-shock magnetic field strength relative to the post-shock sound speed. As with rotation, there are many multi-dimensional effects that complicate the  consideration of magnetic fields as an initiating agent in supernovae. Still, the simple ansatz of Equation~\eqref{eq:ansatz} gives a starting point.
            
            As another example that deserves more critical analysis, several works have considered the importance of wave driving, damping, and propagation in the post-shock region, generated by motions within the proto-neutron star \citep{Burrows2006,Harada2017,Gossan2020}. For a given wave flux into the gain region, the overall wave pressure gradient again enters the Euler equation for the post-shock flow in a manner similar to the turbulent pressure, except with
            \begin{equation}
                \widetilde{R_{ij}} \to 2\overline{P_{ij}^{\rm aco}},
            \end{equation}
            where $P^{\rm aco}$ is the pressure from acoustic waves.
            For monochromatic, adiabatic sounds waves, the ratio of angular to radial  acoustic pressure is
            \begin{equation}
                \alpha = \frac{P_{\theta\theta} + P_{\phi\phi}}{P_{rr}} = 2\frac{\gamma-1}{\gamma+1},
            \end{equation}
            where $\gamma$ is the ratio of the specific heats \citep{Lamers1999}.  Thus the acoustic pressure is strongly radially dominant, only approaching our case of anisotropic turbulence in the limit $\gamma\to\infty$. Ultimately, however, the acoustic term in the antesonic condition should look the same as the turbulent term, i.e.,
            \begin{equation}
                \xi_\mathrm{aco} = \frac{2K_{\rm aco}}{v_\esc^2},\quad a_\mathrm{aco} = \frac{1}{3}.
            \end{equation}
            Here $K_{\rm aco}$ is the specific acoustic kinetic energy:
            \begin{equation}
                K_{\rm aco} = \frac{1}{2}\overline{v_\mathrm{aco}^2} = \frac{1}{4}\omega^2\zeta^2,
            \end{equation}
            where $v_{\rm aco}$ is the velocity associated with acoustic motions, $\omega$ is the frequency of those oscillations, and $\zeta$ is the amplitude of the physical motion associated with the waves, i.e., a particle in the post-shock accretion flow traces a path governed by
            \begin{equation}
                r(t) = \overline{r}(t) + \zeta\cos(\omega t).
            \end{equation}
            
            In recent calculations of acoustically powered explosions, \citet{Harada2017} found that successful explosions generally had lower thermal antesonic ratios (in their case, $\max\left(c_s^2/v_\esc^2\right)$) relative to failed explosions.  In light of our work, this is expected because by increasing the acoustic power, \citet{Harada2017} are lowering the critical $\xi_\mathrm{th}$ below the maximum value of $\xi_\thm$ reached in their non-exploding case.  Specifically, \citet{Harada2017} use waves of frequency $\omega\sim\SI{2}{\per\ms}$ and amplitudes of order $\zeta\sim\;$\numrange{5}{10}\si{\km} near the shock.  In our framework, this would result in an acoustic antesonic ratio of $\xi_\mathrm{aco}\approx\;$\numrange{0.03}{0.007} for a shock radius $r_\mathrm{sh} = \SI{150}{\km}$, and a \numrange{16}{4} per-cent reduction in the critical thermal antesonic ratio $\xi_\thm$ required for explosion.  We note that the actual decrease in the critical neutrino luminosity found by \citet{Harada2017} is within this range.
        
        \subsection{1D  Simulations with Turbulence}\label{sec:couch}
            \citet{Couch2020} (hereafter in this section, CWO20) perform 1D supernova simulations using mixing length theory to approximate turbulence.  The strength of turbulence is parameterized by $\alpha_\Lambda$:
            \begin{equation}
                \alpha_\Lambda = \Lambda_\mathrm{mix}\frac{\rho g}{P},
            \end{equation}
            where $\Lambda_\mathrm{mix}$ is the mixing length.  The value $\alpha_\Lambda=0$ corresponds to a simulation without turbulence, for which explosions are not achieved. As $\alpha_\Lambda$ increases above a certain threshold, progenitors begin to explode at different times, producing a range of explosion energies and neutron star masses. The results can be tuned to produce results for large surveys of progenitors and compared to the works of \cite{Ugliano2012,Pejcha2015,Sukhbold2016}. 
            
            Because in CWO20 the strength of the turbulence is a free parameter, in principle, it should be possible to make an apposite comparison with our results. For small $\alpha_\Lambda$ and no explosion, the critical antesonic ratio should not be reached during the simulation, while models with bigger $\alpha_\Lambda$ just beyond the threshold for explosion should just exceed the critical generalized turbulent antesonic condition identified in Equation~\eqref{eq:turbante}. That is, we expect a given model set with turbulence to explode only once the critical condition is exceeded.
            
            In practice, the comparison is complicated by the fact that we have assumed an isothermal model for the gas, whereas the simulations use a general EOS and include neutrino heating/cooling.  Nevertheless, we can attempt to make the comparison by first noting that a polytropic EOS ($P\propto\rho^\Gamma$) is a better fit to these models than the isothermal results presented in this paper. \citet{Raives2018} show analytically that the thermal antesonic ratio in the polytropic limit becomes (see also \citealt{Pejcha2012})
            \begin{equation}
                \frac{c_s^2}{v_{\rm esc}^2}=\frac{3\Gamma}{16}.\label{eq:antepolythm}
            \end{equation}
            Using this result and the calculations presented here (eq.~\ref{eq:turbante}), we can write down an approximate critical antesonic condition for a polytropic EOS with turbulence. While we are unable to derive a closed form analytic solution to the equations, we find an approximate numerical solution to the critical condition that is a good fit over the range of parameters we have tested:
            \begin{equation}
                \xi_\mathrm{total} = \xi_\thm + \left(\frac{1}{3}+\frac{\Gamma-1}{14.4205}\right)\xi_\tb \approx \frac{3}{16}\Gamma. \label{eq:gammaante}
            \end{equation}
            Here, and throughout this section, $\xi_\thm$ is the polytropic antesonic ratio given by Equation~\eqref{eq:antepolythm}.  Equation~\eqref{eq:gammaante} is most accurate for small $\xi_\tb$.  Below $\xi_\tb\sim0.05$, the error in $\xi_\mathrm{total}$ from our approximate solution is $\lesssim1$ per-cent for all $\Gamma\leq\frac{5}{3}$.  As $\Gamma\to\frac{5}{3}$, (and as $\Gamma\to1$, for which the approximation approaches Equation~\ref{eq:turbanteappr}), a sub per-cent error in the critical condition is maintained for turbulent antesonic ratios as large as $\xi_\tb\simeq0.10-0.15$.
            
           The work of \cite{Pejcha2012} shows that for models with neutrino heating/cooling and a general EOS that the antesonic condition is reached near the gain radius and that the antesonic condition is best represented in such cases by the maximum value achieved in the gain region, rather than just being evaluated at the shock as in the purely isothermal problem. Thus, for the purposes of comparison to CWO20, we take the total antesonic ratio to be
           \begin{equation}
                \xi_\mathrm{total} = \max\left[\xi_\thm + \left(\frac{1}{3}+\frac{\Gamma-1}{14.4205}\right)\xi_\tb\right], \label{eq:gammaante2}
            \end{equation}
            and we calculate this quantity throughout the time evolution of two progenitors, simulated with a range of $\alpha_\Lambda$.
            
\begin{figure}
                \centering
                \includegraphics[width=\linewidth]{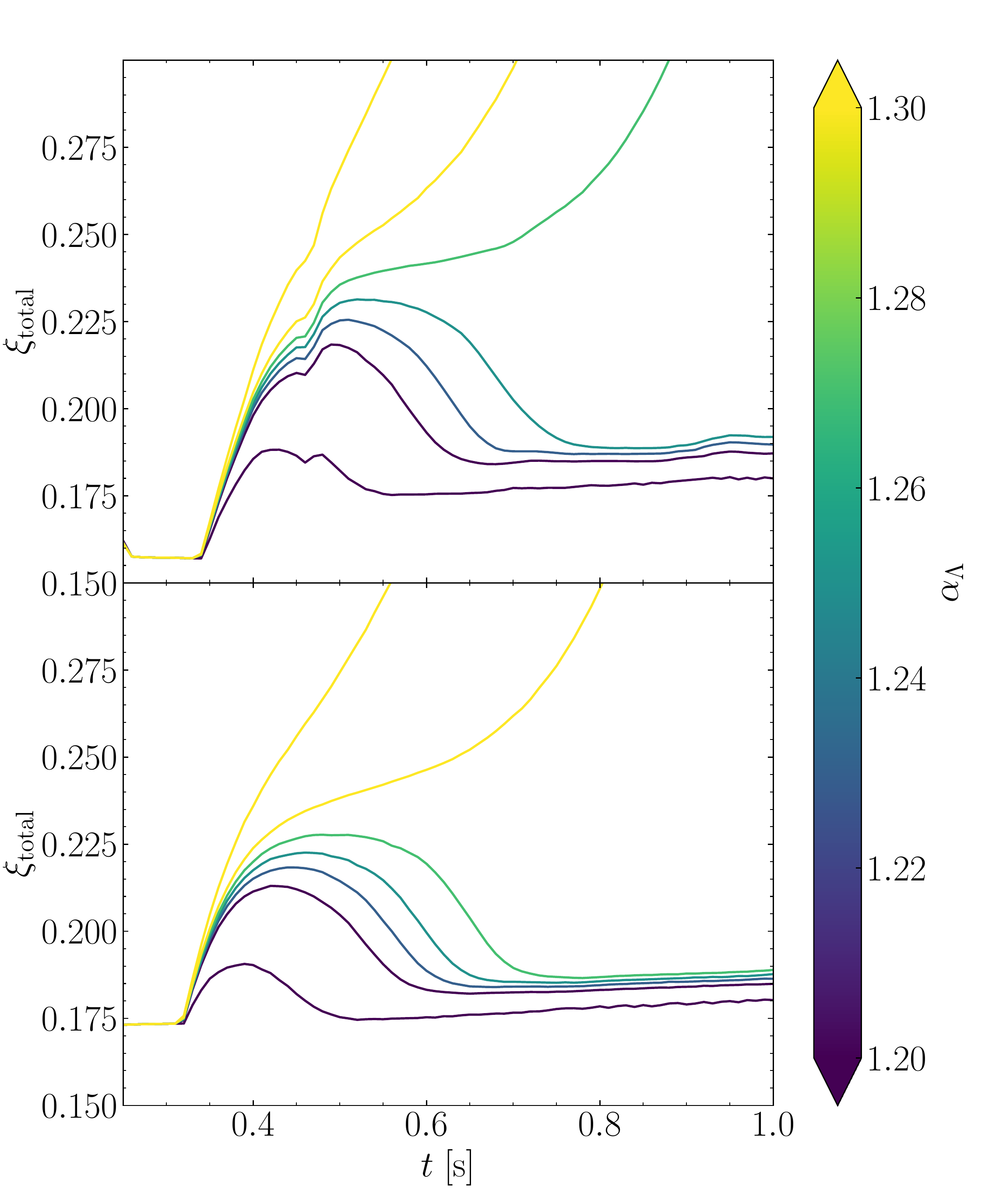}
                \caption{Total antesonic ratio (Equation~\ref{eq:gammaante2}) as a function of time for different values of the turbulent parameter $\alpha_\Lambda$.  The top panel uses a \SI{13}{M_\odot} progenitor and the bottom panel uses a \SI{15}{M_\odot} progenitor.  We see that, by increasing $\alpha_\Lambda$, we can induce explosion in otherwise stable configurations.  Though the transition from accretion to wind solutions occurs at slightly lower $\xi$ than we might expect, this can largely be attributed to differences between the simulation and the "pure" polytropic analysis.}
                \label{fig:turbalpha}
            \end{figure}

            Understanding these caveats and with Equation~\ref{eq:gammaante2} in hand, in Figure \ref{fig:turbalpha} we plot the total antesonic ratio as determined by the CWO20 simulations, for two different progenitor models: a \SI{13}{M_\odot} and \SI{15}{M_\odot} progenitor from the set used by \citet{Sukhbold2016}. The location of the separatrix between accretion and wind (i.e., exploding) solutions occurs for $\xi\gtrsim0.23$. For $\Gamma\simeq\frac{4}{3}$, as implied by the simulations at the maximum of $\xi$ in the profiles, we would instead expect a critical value of $\xi\simeq 3\Gamma/16=\frac{1}{4}$. Put another way, a value of $\Gamma\sim1.23$ would be needed to make Equation~\ref{eq:gammaante} accord with the observed critical value in Figure~\ref{fig:turb}. The $\sim10$ per-cent difference between the expected critical value and that derived from the simulations is likely caused by differences between the physics of the simulations and the ``pure'' polytropic EOS analysis used to derive Equations ~\eqref{eq:gammaante} and \eqref{eq:antepolythm}.  In particular, these simulations include an explicit coupling term between the turbulent and non-turbulent velocities that is not present in our analysis.  Furthermore, these simulations include a general EOS and neutrino heating and cooling terms that are likewise absent from our models. Indeed, the accretion region between the proto-neutron star and the shock is not well-approximated by a single polytropic EOS. A focus of future work could be to make a more direct and sharper  comparison of our results for the generalized antesonic condition with  time-dependent numerical models by using more realistic thermodynamics.
        
        \subsection{Multidimensional Simulations}\label{sec:multidim}
        
            Since the pioneering work of \cite{Herant1994,Burrows1995,Janka1995} evidence has accumulated that breaking spherical symmetry leads core-collapse supernova simulations closer to the condition for explosion. Indeed, many 1D models in the literature fail \citep{Rampp2000,Liebendoerfer2001,Thompson2003}, while their multi-dimensional counterparts sometimes succeed (but, see \citealt{OConnor2018}), albeit with low explosion energies in some cases (e.g., \citealt{Vartanyan2019}). The first study of the \cite{Burrows1993} critical curve in 2D by \cite{Murphy2008} showed that the critical threshold in neutrino luminosity for explosion was lower than in 1D at a given mass accretion rate. That investigation was followed by a number of works on the critical threshold for explosion in multi-dimensional simulations (e.g., \citealt{Nordhaus2010,Couch2013}). 
            
            These works and many others show that the character and strength of the post-shock turbulence has a direct impact on the shock dynamics with time-dependent and progenitor-dependent thresholds for explosion. 
            However, much of the discussion of the critical condition for explosion and the success or failure of any given model is qualitative in the sense that there has so far not been a quantitative theory for how much turbulence `helps' in bringing models to explosion. There are a number important exceptions, including \cite{Mueller2015} (discussed below), the work of \cite{Couch2020} (discussed above), and \cite{Murphy2013,Couch2015a} who argue on the basis of momentum balance at the shock that turbulence leads to larger shock radii, and that this aids explosion (see also \citealt{Couch2014,Radice2016}). 
            
            Here, our extension of the antesonic condition including rotation and turbulence in Equation~\eqref{eq:rotandturb}, and the generalized ``ansatz'' for other forces in Equation~\eqref{eq:ansatz} provides a simple and intuitive quantitative criterion for explosion, albeit for a toy model. Specifically, the threshold for explosion in multi-dimensional simulations is lower because the critical threshold for the thermal energy content of matter in the post-shock region is lower by (Equation~\ref{eq:turbante})
            \begin{equation} \xi_\thm\simeq\frac{3}{16}-\frac{1}{3}\xi_{\rm turb}
            \end{equation}
            (for small $\xi_{\rm turb}$). The physics of this condition is the same as for the antesonic condition itself: above a certain critical value of the linear combination of $\xi_\thm$ and $\xi_{\rm turb}$ it is impossible to simultaneously satisfy the strong shock jump conditions and the time-steady Euler equation and the system undergoes time-dependent transition from accretion to thermal wind. Because the strength and character (i.e., level of isotropy) of the turbulence varies as a function of progenitor and time, the exact values of $\xi_\thm$ and $\xi_{\rm turn}$ will likewise also vary with progenitor and time.
            
            While much work remains in attempting to apply this condition to multi-dimensional, full physics models, this explosion condition is useful as a prior on which models will explode and which will fail, and potentially provides the physical explanation for  the explosion itself. Future works can directly test if an equation like (\ref{eq:turbante}) is indeed the correct way to think about this threshold. 
            
            In this context, ideally we would be able to make a direct connection between the explosion condition of Equation~\eqref{eq:turbante} and the results of multi-dimensional simulations. One example is the work of \citet{Mueller2015} (hereafter in this section, MJ15) who perform 2D simulations with turbulence, and find a threshold in the turbulent Mach number above which all of their simulations explode.  However, the comparison wtih MJ15 is complicated by several factors.  MJ15 define the turbulent Mach number using the angular turbulent velocity rather than the radial, i.e.,
            \begin{equation}        
                \widetilde{\mach_\mathrm{MJ15}^2} = \frac{\widetilde{v_\theta''v_\theta''}}{c_s^2}.
            \end{equation}
            This differs from our definition of the turbulent Mach number by a factor of $1/\alpha$:
            \begin{equation}
                \widetilde{\mach^2_\tb} \sim \frac{1}{\alpha}\widetilde{\mach_\mathrm{MJ15}^2}.
            \end{equation}
            To obtain the specific turbulent kinetic energy $K$, which enters our definition of $\xi_{\rm turb}$ we must further multiply by a factor of $1+\alpha$ to account for both the radial and non-radial motions:
            \begin{equation}
                \frac{1}{2}\frac{\xi_{\rm turb}}{\xi_\thm}=\frac{K}{c^2_T} = \frac{1+\alpha}{2}\widetilde{\mach^2_\tb} = \frac{1+\alpha}{2\alpha}\widetilde{\mach_\mathrm{MJ15}^2}.
            \end{equation}
            
            An additional complication is that to make a direct comparison between MJ15 and this work requires knowledge of the relative strength of radial and non-radial turbulent motions (quantified by $\alpha$) at the time of explosion, which MJ15 do not provide. While MJ15 initialize their simulations with a wide range of $\alpha$ (up to $\alpha\simeq17$), we assume $\alpha$ evolves to equipartition between the radial and non-radial motions (i.e., $\alpha=1$; \citealt{Arnett2009}) by the time of explosion.
            
            Furthermore, while here we have explored a wide, two-dimensional parameter space of $\xi_\thm$ and $\xi_\tb$, for many different steady-state values of the mass accretion rate (e.g., Fig.~\ref{fig:turbAnte}) in MJ15 the post-shock sound speed (and thus, $\xi_\thm$) set by neutrino heating and cooling, the accretion rate, and the shock radius, which is determined by \emph{both} the sound speed and the specific turbulent kinetic energy $K$. Thus, the critical condition found by MJ15 is essentially for fixed $\xi_\thm$. While we cannot determine what $\xi_\thm$ was at explosion in the simulations of MJ15, we can determine what the critical $\xi_\thm$ would be in our model using the critical turbulent Mach number found by MJ15.
            
            MJ15 finds a critical turbulent Mach number 
            \begin{equation}
                \widetilde{\mach_\mathrm{MJ15}^2} \simeq 0.3.
            \end{equation}
            For $\alpha=1$, this corresponds to the same value of $\widetilde{\mach^2_\tb}$.  Using the approximate antesonic condition for the polytropic equation of state we found earlier (Equation~\ref{eq:gammaante}), we would expect the critical thermal antesonic ratio to be
            \begin{equation}
                \xi_\th \approx 0.227
            \end{equation}
            for $\Gamma=\frac{4}{3}$.  This is about a 10 per-cent decrease from the limit of $\frac{3}{16}\Gamma=\frac{1}{4}$. Put another way, if we assume a one-to-one correspondence between the critical neutrino luminosity and the critical $c_s^2$ (i.e., a factor of 2 change in one means a factor of 2 change in the other), then the 25 per-cent change in the critical neutrino luminosity measured by MJ15 would mean a critical total antesonic ratio of
            \begin{equation}
                \xi_\mathrm{total} \approx 0.21,
            \end{equation}
            assuming the same critical turbulent Mach number.  This is about 15 per-cent below what our analytic theory predicts, similar to the discrepancy we found between our theory and the results of \citet{Couch2020} in the previous section.  Likewise, we can attribute this discrepancy to the difference between our simplified use of a ``pure'' polytropic EOS and the physical EOS used by MJ15.  There is also the potential for the dynamical effects of turbulence, in particular, oscillations in the shock surface, to lower the critical curve, as in the work of \cite{Murphy2008} and \cite{Gabay2015}.
        
        \section{Conclusions}
            We demonstrate that both rotation and turbulence reduce the critical antesonic ratio in our 1D, isothermal model.  We emphasize that attempting to characterize a simulation only by the thermal antesonic ratio  can be misleading. In a model with significant rotation or turbulence, we would expect the critical value of $\xi_{\th}$ to be measurably smaller than one would expect from the non-turbulent analysis. 
            
            Specifically, if we consider our first order approximation to the antesonic condition (Equation~\ref{eq:firstorder}), this is equivalent to stating that
            \begin{equation}
                \xi_{\thm,\crit} = \frac{3}{16} - \frac{1}{3}\xi_\cen - \frac{1}{3}\xi_\tb.
            \end{equation}
            As discussed in \S\ref{sec:multidim}, in this picture, multi-dimensional models are easier to explode than their spherical counterparts because the turbulent pressure in the post-shock region and at the shock decreases the critical sound speed required for explosion.  In a comparison like that presented in \citet{Nagakura2019}, where a low-resolution model fails and a high-resolution model succeeds, we would interpret the success and failure in terms of the critical condition above (see also \citealt{Raives2018}, where we discuss the effect of resolution on the critical condition).  As discussed in \S\ref{sec:turbulence} (see Figure~\ref{fig:turbAnte}) and \S\ref{sec:couch}, the turbulent Mach numbers needed to substantially affect the critical condition are typical of the results of multi-dimensional simulations. For example, \citet{Couch2015a} finds ratios of turbulent to thermal pressure of up to $K/c_T^2\sim0.5$; in our model that would correspond to a decrease of the critical curve normalization (i.e,. the critical $c_T^2$ for a given $\dot{M}$) of 15 to 25 per cent (decreasing with larger $\dot{M}$).  For rotation, the effect is less significant, with a rotation rate corresponding to a millisecond period remnant only leading to a 1 to 4 per cent decrease.  Faster rotation rates can lead to more modest decreases in the critical curve normalization, but we quickly reach rotation rates that would lead to remnants rotating much faster than supported by observation, and rates approaching the limits discussed in \S\ref{sec:rotation}.
            
            
            We also explore the potential of an antesonic ``ansatz'' to describe the effects of other forces on the critical condition for explosion.  Our analysis suggests that there is an antesonic ratio (i.e., some characteristic velocity squared divided by the escape velocity squared) associated with each force, and that, to first order, the critical condition can be expressed as a linear combination of these terms.  As examples, we briefly discuss magnetic pressure and acoustic wave pressure in the context of this framework.
            
            In our models, we find theoretical maximum values of $\xi_\cen=\frac{1}{2}$ and $\xi_\tb=\frac{1}{2},\frac{9}{16}$ (for isotropic and anisotropic turbulence, respectively).  Our antesonic ``ansatz'' suggests similar limits exist for other forces.  At these values of the non-thermal antesonic ratio, the pressure behind the shock is large enough to lead to explosion even in the absence of thermal pressure.  However, in nature, we cannot actually reach this hypothetical limit of zero thermal pressure.  Even in the absence of neutrino heating from the PNS core, the accreting material will be shock heated as it passes the shock, providing a minimum thermal pressure and thus minimum thermal antesonic ratio.  Because of this, the actual maximum non-thermal antesonic ratios will be smaller than the limits presented in this paper.  In future work, we will self-consistently consider the effect of shock heating and the limits it imposes on non-thermal contributions to explosion.
            
            However, we stress that an exploration of these effects in full-physics supernova simulations is required before we can make specific numerical predictions about those cases.  In non-isothermal simulations of this problem, the numerical value of the antesonic condition will generally change.  As shown by \cite{Raives2018}, in the case that the post-shock fluid is described by an EOS of the form $P\propto\rho^\Gamma$, the antesonic condition is $c_s^2/v_\esc^2=3\Gamma/16$ at the shock, where $c_s=(\Gamma P/\rho)^{1/2}$ is the adiabatic sound speed.  Similarly, for an equation of state coupled to neutrino heating and cooling, \citet{Pejcha2012} showed that an antesonic condition of $\xi^\nu_\mathrm{crit} \equiv \max (c_s^2/v_\esc^2)\sim0.2$ was a good fit over several orders of magnitude, though later multidimensional simulations have suggested a wider range of critical antesonic ratios, $\xi^\nu_\mathrm{crit}\sim 0.2-0.3$ \citep{Couch2013a,Dolence2013,Couch2014}.
            
            We also note that our results on rotation suggest that rapidly rotating stars should be easier to explode near the equator (where $v_\phi$ is the largest), in contrast to \citet{Yamasaki2005} which prefers polar explosions.  In reality, these are not two contradictory results but two competing effects.  We find that the critical thermal antesonic ratio is smallest at the equator (implying explosion should initiate there), while \citet{Yamasaki2005} find that rotation focuses the accreting material towards the equator, decreasing the accretion rate at the poles (implying explosion should commence along the rotation axis).  Which of these two effects win out will likely depend on details of the equation of state, as well as the relative values of $v_\phi$ $P_\mathrm{th}$, and $\dot{M}(t)$.  E.g., we might na\"ively expect models with extreme accretion rates to care more about the distribution of accretion over the shock surface, while models with very small accretion rates and/or small thermal pressures might care more about the reduction in $\xi_{\th,\crit}$ near the equator.  Regardless, such questions must be answered with 2D and 3D simulations of this problem and must wait, ideally, for a multi-D theory of the antesonic condition to be developed.
    
    \section*{Acknowledgements}
    MJR and TAT thank Davide Radice, Bernhard M\"uller, and Ondrej Pejcha for helpful conversations. TAT acknowledges support from a Simons Foundation Fellowship and an IBM Einstein Fellowship from the Institute for Advanced Study, Princeton.  TAT and MJR also acknowledge partial support from NASA grant 80NSSC20K0531.
    
    SMC is supported by the U.S. Department of Energy, Office of Science, Office of Nuclear Physics, Early Career Research Program under Award Number DE-SC0015904. This material is based upon work supported by the U.S. Department of Energy, Office of Science, Office of Advanced Scientific Computing Research and Office of Nuclear Physics, Scientific Discovery through Advanced Computing (SciDAC) program under Award Number DE-SC0017955. This research was supported by the Exascale Computing Project (17-SC-20-SC), a collaborative effort of the U.S. Department of Energy Office of Science and the National Nuclear Security Administration.
    
    The research presented in this paper made use of the python libraries \textsc{Matplotlib} \citep{Hunter2007}, \textsc{Numpy} \citep{vanderWalt2011}, and \textsc{SciPy} \citep{Virtanen2020}.  It also makes use of \textsc{yt-project} \citep{Turk2011}.
    
    \section*{Data Availability}
    The data underlying this article will be shared on reasonable request to the corresponding author.  Requests for data originating from \citet{Couch2020} should be sent directly to SMC.
    
    
    \newpage
    
    \bibliographystyle{mnras}
    \bibliography{antesonic}
    
    \appendix
    
    \section{Properties of the Favre Average}
        In this section, we review the definition and useful rules of the Favre average.  Recall, for a field $f$, split into a (Favre-averaged) background and turbulent component
        \begin{equation}
            f = \widetilde{f} + f'',
        \end{equation}
        the Favre average is defined as:
        \begin{equation}
            \overline{\rho}\widetilde{f} = {\overline{\rho f}},
        \end{equation}
        where $\overline{f}$ is the Reynolds average
        \begin{equation}
            \overline{f} = \frac{1}{\Delta T}\int_T f(t)\:\mathrm{d}t.
        \end{equation}
        Note that by definition:
        \begin{equation}
            \widetilde{f''} = 0.
        \end{equation}
        The Favre average is thus a density-weighted Reynolds average.  Ultimately, the Favre average is an integration operator, and thus the normal rules for integration imply the following identities, which we make use of in the next section:
        \begin{align}
            \overline{\rho}\widetilde{\left(f+g\right)} &= \overline{\rho}\left(\widetilde{f} + \widetilde{g}\right)\\
            \overline{\rho}\left(\widetilde{g\cdot f''}\right) &= 0\\
            \overline{\rho}\skew{5.5}\widetilde{\widetilde{f}} &= \overline{\rho}\widetilde{f}\\
            \overline{\rho}\widetilde{\left(\widetilde{f}\cdot g\right)} &= \overline{\rho}\left(\widetilde{f}\cdot\widetilde{g}\right)\\
            \widetilde{\frac{\partial f}{\partial s}} &= \frac{\partial\widetilde{f}}{\partial s}.
        \end{align}
    
    \section{Favre-Averaged Euler Equations}\label{app:Favre}
        We start with the full, time-steady Euler equations, presented here in vector form:
        \begin{align}
            \nabla\cdot(\rho\mathbf{v}) &= 0\\
            \nabla\cdot(\rho\mathbf{v\otimes v}) + \nabla{P} - \rho\mathbf{g} &= 0
        \end{align}
        We then take decompose $\mathbf{v}$ into its background component $\widetilde{\mathbf{v}}$ and its turbulent component $\mathbf{v''}$:
        \begin{align}
            \nabla\cdot(\rho(\widetilde{\mathbf{v}}+\mathbf{v''})) &= 0\\
            \nabla\cdot(\rho(\widetilde{\mathbf{v}}+\mathbf{v''})\otimes(\widetilde{\mathbf{v}}+\mathbf{v''})) + \nabla{P} - \rho\mathbf{g} &= 0
        \end{align}
        Then we take a Reynolds average of the full equation:
        \begin{align}
            \overline{\nabla\cdot(\rho(\widetilde{\mathbf{v}}+\mathbf{v''}))} &= 0\\
            \overline{\nabla\cdot(\rho(\widetilde{\mathbf{v}}+\mathbf{v''})\otimes(\widetilde{\mathbf{v}}+\mathbf{v''})) + \nabla{P} - \rho\mathbf{g}} &= 0
        \end{align}
        Using the rules outlined in the previous section:
        \begin{align}
            \nonumber\nabla\cdot(\overline{\rho(\widetilde{\mathbf{v}}+\mathbf{v''})}) &= 0\\
            \nonumber\nabla\cdot(\overline{\rho\widetilde{\mathbf{v}}}+\cancel{\overline{\rho\mathbf{v''})}}) &= 0\\
            \nabla\cdot(\overline{\rho}\widetilde{\mathbf{v}}) &= 0\label{eq:ContinuityFavre}
        \end{align}
        And for the momentum equation:
        \begin{align}
            \nonumber\nabla\cdot(\overline{\rho(\widetilde{\mathbf{v}}+\mathbf{v''})\otimes(\widetilde{\mathbf{v}}+\mathbf{v''})}) + \nabla{\overline{P}} - \overline{\rho}\mathbf{g} &= 0\\
            \nonumber\nabla\cdot(\overline{\rho\widetilde{\mathbf{v}}\otimes\widetilde{\mathbf{v}}} + \cancel{2\overline{\rho\widetilde{\mathbf{v}}\otimes\mathbf{v''}}} + \overline{\rho\mathbf{v''}\otimes\mathbf{v''}}) + \nabla{\overline{P}} - \overline{\rho}\mathbf{g} &= 0\\
            \nonumber\nabla\cdot(\overline{\rho}\widetilde{\mathbf{v}}\otimes\widetilde{\mathbf{v}}) +  \nabla\cdot\widetilde{\mathbf{R}} + \nabla{\overline{P}} - \overline{\rho}\mathbf{g} &= 0.
        \end{align}
        The divergence term we will transform as
        \begin{equation}
            \nabla\cdot(\overline{\rho}\widetilde{\mathbf{v}}\otimes\widetilde{\mathbf{v}}) = \widetilde{\mathbf{v}}\cancel{(\nabla\cdot(\overline{\rho}\widetilde{\mathbf{v}}))} + \overline{\rho}\widetilde{\mathbf{v}}\cdot\nabla\widetilde{\mathbf{v}},
        \end{equation}
        and thus we have
        \begin{equation}
            \overline{\rho}\widetilde{\mathbf{v}}\cdot\nabla\widetilde{\mathbf{v}} +  \nabla\cdot\widetilde{\mathbf{R}} + \nabla{\overline{P}} - \overline{\rho}\mathbf{g} = 0.\label{eq:MomentumFavre}
        \end{equation}
        Here $\widetilde{\mathbf{R}}$ is the Reynolds tensor, the components of which are:
        \begin{equation}
            \widetilde{R_{ij}} = \overline{\rho}\widetilde{v''_iv''_j}
        \end{equation}
        
        Expressing Equations~\eqref{eq:ContinuityFavre} and \eqref{eq:MomentumFavre} in spherical coordinates, we obtain
        \begin{align}
            \frac{1}{\overline{\rho}}\frac{\partial\overline{\rho}}{\partial r} + \frac{1}{\widetilde{v_r}}\frac{\partial\widetilde{v_r}}{\partial r} + \frac{2}{r} &= 0\\
            \overline{\rho}\widetilde{v_r}\frac{\partial\widetilde{v_r}}{\partial r} - \nonumber\overline{\rho}\frac{\widetilde{v_\phi}\widetilde{v_\phi}}{r} + \frac{\partial(\overline{P} + \widetilde{R_{rr}})}{\partial r} &\\
            + (2-\alpha)\frac{\widetilde{R_{rr}}}{r} + \overline{\rho}\frac{GM}{r^2} &= 0,
        \end{align}
        where $\alpha$ is the degree of isotropy of the turbulence, as defined in Equation~\eqref{eq:alpha}.  
        
        We can also write this equation in terms of the turbulent kinetic energy
        \begin{equation}
            K = \frac{1}{2}\sum_i\widetilde{v''_iv''_i}.
        \end{equation}
        Assuming $\frac{\partial K}{\partial r} = 0$ and an isothermal equation of state, we have
        \begin{align}
            \nonumber\widetilde{v_r}\frac{\partial\widetilde{v_r}}{\partial r} - \frac{\widetilde{v_\phi}\widetilde{v_\phi}}{r} + \frac{1}{\overline{\rho}}\left(c_T^2 + \frac{2K}{1+\alpha}\right)\frac{\partial\overline{\rho}}{\partial r} &\\
            + \frac{2-\alpha}{1+\alpha}\frac{2K}{r} + \frac{GM}{r^2} &= 0.
        \end{align}
        Let $c_\eff^2 = c_T^2 + \frac{2K}{1+\alpha}$.  Then:
        \begin{equation}
            \widetilde{v_r}\frac{\partial\widetilde{v_r}}{\partial r} - \frac{\widetilde{v_\phi}\widetilde{v_\phi}}{r} + \frac{c_\eff^2}{\overline{\rho}}\frac{\partial\overline{\rho}}{\partial r} + \frac{2-\alpha}{1+\alpha}\frac{2K}{r} + \frac{GM}{r^2} = 0.
        \end{equation}
        Substituting in Equation~\eqref{eq:ContinuityFavre},
        \begin{equation}
            \left(\widetilde{v_r} - \frac{c_\eff^2}{\widetilde{v_r}}\right)\frac{\partial\widetilde{v_r}}{\partial r} - \frac{\widetilde{v_\phi}\widetilde{v_\phi}}{r} - \frac{2c_\eff^2}{r} + \frac{2-\alpha}{1+\alpha}\frac{2K}{r} + \frac{GM}{r^2} = 0.
        \end{equation}
        Let $\xi_\cen = \widetilde{v_\phi}\widetilde{v_\phi}/v_\esc^2$ and $\xi_\tb = 2K/v_\esc^2$.  Then:
        \begin{align}
            \frac{\widetilde{v_\phi}\widetilde{v_\phi}}{r} &= \frac{v_\esc^2}{r}\frac{\widetilde{v_\phi}\widetilde{v_\phi}}{v_\esc^2} = \frac{2GM}{r^2}\xi_\cen\\
            \frac{2-\alpha}{1+\alpha}\frac{2K}{r} &= \frac{2-\alpha}{1+\alpha}\frac{v_\esc^2}{r}\frac{2K}{v_\esc^2} = \frac{2-\alpha}{1+\alpha}\frac{2GM}{r^2}\xi_\tb,
        \end{align}
        and we have:
        \begin{align}
            \left(\widetilde{v_r} - \frac{c_\eff^2}{\widetilde{v_r}}\right)\frac{\partial\widetilde{v_r}}{\partial r} - \frac{2c_\eff^2}{r} + \frac{GM}{r^2}\Xi_2 = 0\\
            \Xi_2 = 1 - 2\xi_\cen + 2\frac{2-\alpha}{1+\alpha}\xi_\tb
        \end{align}
        Multiplying all terms by ${2GM}/{c_\eff^4}$, and allowing $\widetilde{\mach} = \widetilde{v_r}/c_\eff$:
        \begin{equation}
            \left(\widetilde{\mach} - \widetilde{\mach}^{-1}\right)\frac{\partial\widetilde{\mach}}{\partial \xi_\eff} - \frac{2}{\xi_\eff} + \frac{\Xi_2}{2\xi_\eff^2} = 0,\label{eq:FavreEulerDimless}
        \end{equation}
        where
        \begin{equation}
            \xi_\eff = \frac{c_\eff^2}{v_\esc^2}
        \end{equation}{}
        
    \section{Favre-Averaged Shock-Jump Conditions}\label{app:shockjump}
        We start with the continuity and momentum shock-jump conditions:
        \begin{align}
            \rho_1v_1 &= \rho_2v_2\\
            \rho_1v_1^2 + P_1 &= \rho_2v_2^2 + P_2,
        \end{align}
        Where $v=v_r$ is the radial velocity.  As before, we take the Reynolds average of both equations:
        \begin{align}
            \nonumber\overline{\rho_1v_1} &= \overline{\rho_2 v_2}\\
            \overline{\rho_1}\widetilde{v_1} &= \overline{\rho_2}\widetilde{v_2}
        \end{align}
        \begin{align}
            \nonumber\overline{\rho_1v_1^2 + P_1} &= \overline{\rho_2v_2^2 + P_2}\\
            \nonumber\overline{\rho_1}\widetilde{v_1}\widetilde{v_1} + \overline{\rho_1}\widetilde{v_1''v_1''} + \overline{P_1} &= \overline{\rho_2}\widetilde{v_2}\widetilde{v_2} + \overline{\rho_2}\widetilde{v_2''v_2''} + \overline{P_2}\\
            \overline{\rho_1}\widetilde{v_1}\widetilde{v_1} + \widetilde{R_1} + \overline{P_1} &= \overline{\rho_2}\widetilde{v_2}\widetilde{v_2} + \widetilde{R_2} + \overline{P_2},
        \end{align}
        where $R=R_{rr}$ is the radial Reynolds stress.  Assuming there is no turbulence above the shock, these reduce to
        \begin{align}
            \overline{\rho_1}\widetilde{v_1} &= \rho_2v_2\\
            \overline{\rho_1}\widetilde{v_1}\widetilde{v_1} + \widetilde{R_1} + \overline{P_1} &= \rho_2v_2^2 + P_2.
        \end{align}
        Assuming an isothermal EOS and pressureless free-fall above the shock:
        \begin{align}
            \overline{\rho_1}\widetilde{v_1} &= -\rho_2\sqrt{\frac{2GM}{r}}\\
            \nonumber\overline{\rho_1}\widetilde{v_1}\widetilde{v_1} + \overline{\rho_1}\left(c_T^2 + \frac{2K}{1+\alpha}\right) &= \rho_2\frac{2GM}{r}\\
            \overline{\rho_1}\widetilde{v_1}\widetilde{v_1} + \overline{\rho_1}c_\eff^2 &= \rho_2\frac{2GM}{r}.
        \end{align}
        From here the solution proceeds analogously to the fully non-turbulant case; i.e.,
        \begin{equation}
            \widetilde{\mach_1} = \frac{1}{2}\left(-\xi_\eff^{-1/2} + \sqrt{\xi_\eff^{-1} - 4}\right)\label{eq:appFavreMach}
        \end{equation}
        
    \section{Generalized Favre-Averaged Antesonic Condition}
    Using Equation~\eqref{eq:appFavreMach}, we can now solve Equation~\eqref{eq:FavreEulerDimless} for the antesonic condition:
    \begin{equation}
        \xi_\eff = \frac{1}{32}\left(8\Xi_2 + \sqrt{25 - 16\Xi_2} - 5\right),
    \end{equation}
    which, in the limit $\xi_\cen\ll1,\;\xi_\tb\ll1$, reduces to
    \begin{equation}
        \xi_\thm + \frac{1}{3}\xi_\cen + \frac{1}{3}\xi_\tb = \frac{3}{16}.
    \end{equation}
    
\bsp
\label{lastpage}
\end{document}